\documentclass[11pt,nofootinbib]{article}
\pdfoutput=1
\usepackage{jheppub}
\usepackage[force]{feynmp-auto} 
\usepackage{amsmath}    
\usepackage{graphicx}   
\usepackage{subcaption} 
\usepackage{verbatim}   
\usepackage{color}      
\usepackage{hyperref}   
\usepackage{slashed}
\usepackage{adjustbox}  

\newcommand{\be}{\begin{equation}}
\newcommand{\ee}{\end{equation}}
\newcommand{\bea}{\begin{eqnarray}}
\newcommand{\eea}{\end{eqnarray}}
\newcommand{\nn}{\nonumber}

\begin{document}

\title{Searching for light new physics at the LHC via lepton-number violation }
\author[a]{Wafia Bensalem,}
\author[b]{David London,} 
\author[a]{Daniel Stolarski,}
\author[a]{Alberto Tonero} 
\affiliation[a]{Ottawa-Carleton, Institute for Physics, Carleton University
1125 Colonel By Drive, Ottawa, ON, K1S 5B6, Canada}
\affiliation[b]{Physique des Particules, Universit\'e de Montr\'eal, 1375 Avenue Th\'er\`ese-Lavoie-Roux, Montr\'eal, QC, Canada  H2V 0B3; \\ email: london@lps.umontreal.ca}

\abstract{
We study the collider phenomenology of a simplified model containing a right-handed $W$ in which the $W_R$ couples predominantly to the third generation in the quark sector. The model also includes a light Majorana neutrino,  with $M_1\sim {\cal O}(100)$ GeV, giving rise to lepton-number-violating signatures that are visible at the LHC. Taking into account all the searches from the LHC and Tevatron, we find that this $W_R$ can still be as light as $M_R \sim 300$ GeV. We show that this type of new physics, and others like it, can be detected at the LHC using final states with three same-sign same-flavour leptons.}

\preprint{
{\flushright
UdeM-GPP-TH-21-292 \\
}}

\maketitle

\section{Introduction}

The standard model (SM) of particle physics has made a multitude of predictions, almost all of which have been verified. There is no doubt that it is correct. However, it does not explain a number of other phenomena, such as dark matter, the baryon asymmetry of the universe, neutrino masses, etc. It is therefore incomplete: there must be physics beyond the SM. The LHC has now taken a significant amount of data at the TeV scale, but as yet has found no evidence for new physics (NP). For this reason, there has been significant activity in parameterizing the NP via an effective field theory (EFT). Given the energy of collisions at the LHC, the EFT paradigm is only strictly applicable if the mass scale of new physics is several TeV or larger. Current bounds on EFT operators extracted from LHC data are typically at the TeV scale \cite{Ethier:2021bye,Ellis:2020unq}.  

Still, although the searches and measurements performed at the LHC cover a tremendous range of NP theories, it cannot be excluded that NP exists well below the TeV scale, but has evaded detection. Indeed, it is imperative to continue to check this possibility. To this end, it is useful for theorists to examine types of NP that are light, but still respect all the constraints from LHC. More importantly, it is necessary to find ways of detecting this NP. 

With this in mind, in this paper we consider a simplified model with a right-handed (RH) $W$ in which the $W_R$ couples predominantly to the third generation in the quark sector. Suppressing couplings to quarks in the first two generations significantly eases bounds, mostly because production rates at hadron colliders can be orders of magnitude smaller. This idea has been exploited in variours other contexts, for example in Refs.~\cite{Kamenik:2018nxv,Hayreter:2019dzc,Gintner:2019fvr,Stolarski:2020cvf}. As we will see, the $W_R$ in the simplified model here can be as light as $\sim$ 300 GeV. 
The question then is: how can we detect such a $W_R$?

Our simplified $W_R$ model also contains three RH neutrinos. Because these RH neutrinos are singlets under all SM gauge groups, a Majorana mass for theses states is expected. Our simplified model assumes that one of the RH neutrinos, $N_1$ is light and can be produced in on-shell $W_R$ decays. Because this $N_1$ state is a Majorana fermion,   
it can decay to $\ell^\pm W_R^{*\mp}$: it carries no conserved charge so it can decay to a charged lepton of either sign. As we will show, this allows this type of light NP to be detected through the observation of lepton-number-violating (LNV) signals. Looking for lepton-number-violating production of Majorana neutrinos is an old idea~\cite{Keung:1983uu,Datta:1993nm,Ali:2001gsa,Han:2006ip,Dev:2013wba}, and there have also been experimental searches for such states~\cite{ATLAS:2012ak,CMS:2015qur,CMS:2016aro,CMS:2018iaf,ATLAS:2019kpx,LHCb:2020wxx}. These analyses assume that the $N$ can couple to the SM $W_L$ boson or to a $W_R$ that couples universally to the quarks, but this need not be the case. Furthermore, a  coupling to the $W_L$ can induce mixing between the $N$ and the SM neutrinos, which is strongly constrained~\cite{Deppisch:2015qwa}. Therefore, in order to explore novel phenomenology, we assume that the coupling of the $N_i$ to the $W_L$ and mixing with the $\nu_L$ are both absent, and that the $N_i$ only couples to the $W_R$ whose quark couplings are dominantly to the third generation, as discussed above.

In this work we perform a detailed exploration of the bounds on this simplified model, focusing on the regime where the $W_R$ is relatively light. After enumerating the possible final states, we find that the strongest bounds on the model come from an ATLAS search for same-sign dileptons performed at 7 TeV~\cite{ATLAS:2012ak}. Some bounds can be extracted from more recent searches, but most analyses at higher energies focus on higher-mass $W_R$s and leave the region $M_R \sim 300 - 500$ GeV still unexplored. 

We also study the discovery potential at the LHC of this model, proposing a same-sign three-lepton (SS3L) search. Such a final state has been considered in the phenomenology literature~\cite{Ozcan:2009qm, Mukhopadhyaya:2010qf, Bambhaniya:2013yca, Guo:2017ybk, Agarwalla:2018xpc,Mondal:2021bxa} in the context of other models, some of which are lepton-number conserving, but as far as we are aware, there has been no experimental search for this signature. We do a detailed analysis of the background, taking into account both fake leptons and charge mismeasurement, and map out the discovery potential of the LHC using conservative background assumptions. For concreteness, we focus on the scenario with three same-sign electrons, and we briefly comment on the other flavour possibilities in our conclusions. 

We must stress that we are not advocating this particular model of NP. Rather, it is presented as a sort of ``existence proof'' that some models with light NP are still viable. And the LNV signal of NP, which can be searched for at the LHC, could be generated in other NP models.

We begin in Sec.~2 with a description of our simplified $W_R$ model. We discuss the production and decay of the $W_R$ in Sec.~3, and we establish the present lower limit on $M_R$ in Sec.~4. In Sec.~5, we identify the LNV signal for potential discovery and examine the prospects for its measurement at the LHC, taking into account the SM background. We conclude in Sec.~7, and the connection with the $\nu$SMEFT~\cite{delAguila:2008ir,Aparici:2009fh,Bhattacharya:2015vja,Liao:2016qyd} is made in the Appendix.

\section{\boldmath Simplified $W_R$ model}
\label{sec:model}

We consider a simplified model in which we add a new charged gauge boson $W_R$ that couples to right-handed SM fermions. In the quark sector, we assume that the gauge boson couples dominantly to the third generation. This makes the constraints significantly weaker than for a $W'$ that has significant coupling to the first generation of quarks. As we will see below, this $W_R$ can be as light as $\sim$300 GeV, well below the multi-TeV bound on the traditional $W_R$. 

In the lepton sector, we add three RH neutrinos $N_i$ to fill out the $SU(2)_R$ representation; these are neutral under the SM gauge group. In order to simplify the analysis, we assume that one of the RH neutrinos is light while the other two are heavy and can be ignored. We also assume that the light $N_1$ couples dominantly to a single flavour of SM leptons, which for concreteness we take to be the electron. 
In a full model there will typically also be a neutral gauge boson $Z'$ and scalars to give mass to the $W_R$. We assume that these other states are heavier than the $W_R$ and treat this simplified model without them. 

To set the notation, the Lagrangian is given by 

\begin{eqnarray}
{\cal L} & = & {\cal L}_{{\rm SM}}-\frac{1}{4}W_R^{\mu\nu}W_{R\mu\nu}+M_R^2W_R^{+\mu}W_{R\mu}^{-} +i\bar{N}_{Ri}\slashed\partial N_{Ri} -\frac{1}{2}M_{i}\bar{N}_{Ri}^{c}N_{Ri}+{\rm h.c.}\nonumber \\
 &  & -\frac{ k_{l} \,g }{\sqrt{2}}
 \left[ \bar{N}_{R1}\gamma^{\mu}e_{R}W_{R\mu}^{+}+\bar{N}_{R2}\gamma^{\mu}\mu_{R}W_{R\mu}^{+}+\bar{N}_{R3}\gamma^{\mu}\tau_{R}W_{R\mu}^{+}+{\rm h.c.}\right]\nonumber \\
 &  & -\frac{ k_{q}\, g }{\sqrt{2}}\left[\lambda^{2}\,\bar{u}_{R}\gamma^{\mu}b_{R}W_{R\mu}^{+}+\lambda\,\bar{c}_{R}\gamma^{\mu}b_{R}W_{R\mu}^{+}+\bar{t}_{R}\gamma^{\mu}b_{R}W_{R\mu}^{+}+{\rm h.c.}\right] + ... ~,
\label{Lagrangian}
\end{eqnarray}
where $M_R$ is the $W_{R}$ mass, $M_{i}$ is the Majorana mass of the RH neutrino $N_i$, and we take $M_1 \ll M_{2,3}$. The parameters $k_{l}$ and $k_{q}$ are the sizes of the coupling of leptons and quarks to the $W_R$, relative to the size of the SM $SU(2)$ gauge coupling.  In the quark sector, the only unsuppressed coupling is to the third generation, $tb$. Due to flavour rotations, we do expect couplings to first- and second-generation quarks, but these will be smaller. We parameterize the coupling to $cb$ by $\lambda$ and $ub$ by $\lambda^2$, assuming $\lambda \ll 1$. Couplings to $ts$ and $tu$ are omitted as they are not phenomenologically relevant if they are small. Couplings between only first- and second-generation quarks only are expected to be further suppressed and are also omitted. We take as a benchmark $\lambda = V_{cb} \approx 0.042$, and $k_q = k_l = 1$. Mixing between the $N_i$ and the SM neutrinos $\nu_i$ is very strongly constrained \cite{Deppisch:2015qwa}, so we take it to be absent in our simplified model. 

In Eq.~(\ref{Lagrangian}) the ``$...$'' represents gauge-boson self interactions. The only one that will affect the phenomenology is the coupling of $W_R$ to the photon, whose structure is set by electromagnetic gauge invariance. Specifically, the $ W_{R\mu}^-(p_1)W_{R\nu}^+(p_2)A_\rho(p_3)$-interaction term is given by the following Feynman rule:
\be 
ie[g_{\mu\nu}(p_1-p_2)_\rho+g_{\nu\rho}(p_2-p_3)_\mu+g_{\rho\mu}(p_3-p_1)_\nu] ~,
\ee
where all the momenta $p_i$ are taken to be incoming.

We take the $N_1$ to be lighter than the $W_R$.
In this case, the dominant decay of the $N_1$ is through a virtual $W_R$, but because the $N_1$ is a Majorana state, it can decay to either the positive or negative $W_R$: $N_1 \rightarrow e^\pm (W_R^\mp)^* \rightarrow e^\pm (\bar{q}q')^\mp$. If the $N_1$ is produced in association with a charged lepton, this is the source of the lepton-number violation in this model that will be discussed at length below. 

If kinematically allowed with $M_1 \gtrsim m_t + m_b$, the $(\bar{q}q')^\mp$ will be $tb$, but if not, the next most likely final state will be $cb$. In this work we focus on the case where the $N_1$ decays promptly at colliders, but the decay to $tb$ is kinematically forbidden. This means that we consider 
\begin{equation}
    10 \; \text{GeV} \lesssim M_1 \lesssim  170 \; \text{GeV} ~.
    \label{eq:NDecay}
\end{equation}
Using the results of the tau leptonic decay \cite{Ferroglia:2013dga}, we find that the width is given by
\begin{equation}
\label{Ndecay}
    \Gamma(N_1 \rightarrow e^\pm (W_R^\mp)^* \rightarrow e^\pm (\bar{c}b)^\mp) \simeq 
    \frac{\left(k_l k_q \lambda \right)^2 g^4}{6144\pi^3} \frac{M_1^5}{M_R^4} \ ,
\end{equation}
assuming $M_R \gg M_1 \gg m_b$, and corrections can be included if one of these assumptions is violated~\cite{Ferroglia:2013dga}.

We note, however, that the case of $N$ decay with displaced vertices at colliders can also be of interest (see e.g. \cite{ATLAS:2020wjh,CMS:2021kdm}). Parameterizing a detector as one that can detect a displaced vertex of $x$ millimeters and can detect decays within a detector of size $D$ meters, then a displaced vertex occurs for
\be
\frac{0.003}{D}\leq k_l^2k_q^2g^4\left(\frac{\lambda}{0.042} \right)^2 \left(\frac{M_1}{10\, \text{GeV}} \right)^5\left(\frac{300\, \text{GeV}}{M_R} \right)^4\leq \frac{0.86}{x} \ .
\ee
which can be relevant for very light $M_1$ or very small couplings. Our analysis will work in the regime where the decays are prompt.
\section{\boldmath $W_R$ production and decay}  
In generic left-right-symmetric models, in which the RH charged gauge boson couples universally to all quarks, single resonant production, $pp\to W_R$, is the main channel for $W_R$ production. This process is dominated by light quark-antiquark fusion, in complete analogy to single $W$-boson production in the SM. In the case where the $W_R$ can decay to $q{\bar q}'$, dijet resonance searches are used to put strong constraints on its mass \cite{CMS:2018mgb,CMS:2019gwf}. Another important decay channel is $W_R\to lN$, which is used to search for RH Majorana neutrinos and $W_R$ gauge bosons through lepton-number-violating final states \cite{CMS:2018jxx,ATLAS:2018dcj,CMS:2021dzb}. 

In our model, the $W_R$ does not couple universally to quarks.  Single resonant production of the $W_R$ at a hadron collider proceeds via the $cb$ coupling which is suppressed by $\lambda$ or the $ub$ coupling which is suppressed by $\lambda^2$. Therefore, resonant production can be made  arbitrarily small by reducing the value of $\lambda$.  Other production channels are $pp\to tW_{R}$, $pp\to tbW_{R}$ and $pp\to W_R^+W_{R}^-$. In order to get an idea of the relative importance of the various channels, we take $M_R=300$ GeV, $M_1=100$ GeV and $\lambda=0.042$, and
compute the production cross sections at leading order (LO) for $\sqrt{s} = 13$ TeV using \texttt{MadGraph5\_aMC@NLO} \cite{Alwall:2014hca}. The results are shown in Table~\ref{wrprod}. We see that associated production via $pp\to tW_{R}$ and $pp\to tbW_{R}$ dominates. Pair production is independent of $\lambda$ but is suppressed by the electromagnetic coupling.

\begin{table}[ht]
\centering
\begin{tabular}{|c|c|}
\hline 
process & $\sigma$ [pb]\tabularnewline
\hline 
\hline 
$pp\to tW_R^-$ & 1.564\tabularnewline
\hline 
$pp\to t\bar b W_R^-$ & 0.5874\tabularnewline
\hline 
$pp\to W_R^+W_R^-$ & 0.1965\tabularnewline
\hline 
$pp\to W_R^+$ & 0.1253\tabularnewline
\hline 
\end{tabular}
\caption{LO production cross sections of $W_R$, computed for $M_R=300$ GeV, $M_1=100$ GeV and $\lambda=0.042$.}
\label{wrprod}
\end{table}

The main hadronic decay channel is $W_{R}\to tb$\footnote{$W_R$ lighter than the top is excluded by the crossed process $t \rightarrow W_R b$~\cite{Kamenik:2018nxv}.}; similar to resonant production, the decay width to light quarks, $W_{R}\to cb$ and $W_R\to ub$, are suppressed by $\lambda^2$ and $\lambda^4$, respectively.
There is also a leptonic decay channel, $W_R\to eN_{1}$. Using this same $W_R$-$e$-$N_1$ coupling, the $N_1$ can then decay, but only if $\lambda$ is nonzero (even if small) as shown in Eq.~\ref{eq:NDecay}. This leads to the three-body decay $N_1\to ebc$ (this channel dominates over $N_1\to ebu$ by a factor $1/\lambda^2$). If $\lambda$ is very small or zero, the four (or five) body decay $N_1\rightarrow eb t^* \rightarrow e b b W^{(*)}$ will then be the leading decay of the $W_R$, but we will assume that $\lambda$ is sufficintly large that the three body decay dominates. 

Because the $N$ is a Majorana particle, the decays to $e^- {\bar b} c$ and $e^+ b {\bar c}$ are equally likely. Thus, half the time, the $e$ in $N_1\to ebc$ will have the same charge as the $e$ in $W_R\to eN_{1}$. In this case, the final state includes two same-sign electrons (in this paper, ``electrons'' refers to electrons or positrons). This is a lepton-number-violating final state, and is a smoking-gun signal of NP.

We identify four main NP processes that can produce LNV final states:
\begin{enumerate}
    
\item The LNV0 process $pp\to eN_{1}$ is dominated by single $W_{R}$ production: $pp\to W_{R},\:W_{R}\to eN_{1}$. As noted above, this process can be made arbitrarly small for $\lambda \to 0$. 

\item The LNV1 process $pp\to teN_{1}$ is dominated by $tW_{R}$ associated production: $pp\to tW_{R},\:W_{R}\to eN_{1}$. 

\item The LNV2 process $pp\to tbeN_{1}$ is dominated by $W_{R}$ associated production: $pp\to tbW_{R},\:W_{R}\to eN_{1}$. 

\item The LNV3 process $pp\to eN_{1}eN_{1}$ is dominated by double resonant production: $pp\to W_R W_{R},\:W_{R}\to eN_{1}$. 

\end{enumerate}
The processes LNV0, LNV1 and LNV2 can give final states with two same-sign electrons in association with jets. These are used to produce the strongest constraints on our model (Sec.~\ref{sec:SS2L}). In the LNV3 process, the two $N_1$s can decay to opposite-sign or same-sign electrons. The opposite-sign case is used to place further constraints on our model (Sec.~\ref{sec:multilepton}). On the other hand, the case of same-sign electrons constitutes the novel signal of our model, as it produces a final state with three same-sign electrons (Sec.~\ref{sec:results}). Note that several of these processes have the same diagram topology as the exotic top decay process considered in~\cite{Liu:2019qfa}, but the contribution of these decays is subdominant to the total rate of these final states.

We also identify three main NP processes that give lepton-number-conserving (LNC) final states (i.e., they contain no leptons): 
\begin{enumerate}

\item The LNC0 process $pp\to tb$ is dominated by $W_{R}$ resonant production: $pp\to W_{R},\:W_{R}\to tb$. As always, this process can be made arbitrarily small for $\lambda\to0$. This process is used to constrain the mass of $W_{R}$. 

\item The LNC1 process $pp\to ttb$ is dominated by $tW_{R}$ associated production: $pp\to tW_{R},\:W_{R}\to tb$. 

\item The LNC2 process $pp\to tbtb$ is dominated by $W_{R}$ associated production: $pp\to tbW_{R},\:W_{R}\to tb$. This process is used to set further constraints with the help of associated charged-resonance-production searches (Secs.~\ref{sec:tb} and~\ref{sec:tbTev}). 

\end{enumerate}

The final states produced by all of these LNV and LNC processes are summarized in Table \ref{LNVLNCfinalstates}.

\begin{table}[ht]
\centering
\begin{tabular}{|c|c|}
\hline 
process & final state\tabularnewline
\hline 
\hline 
LNV0 & $e^{\pm}e^{\pm}bj$\tabularnewline
\hline 
LNV1 & $e^{\pm}e^{\pm}bbjjj$\tabularnewline
\hline 
LNV2 & $e^{\pm}e^{\pm}bbbjjj$\tabularnewline
\hline 
LNV3 & $e^{+}e^{-}e^{\pm}e^{\pm}bbjj$\tabularnewline
\hline 
LNC0 & $bbjj$\tabularnewline
\hline 
LNC1 & $bbbjjjj$\tabularnewline
\hline 
LNC2 & $bbbbjjjj$\tabularnewline
\hline 
\end{tabular}
\caption{Final states produced by the various LNV and LNC NP processes.}
\label{LNVLNCfinalstates}
\end{table}

\section{\boldmath Current bounds on the $W_R$} 
\label{sec:constraints}

Experiments have not performed dedicated searches for this particular $W_R$. Instead they have looked at a variety of final states, looking for deviations from the predictions of the SM. We have performed a thorough search of the literature, translating the published measurements into bounds on the $W_R$. In this section, we describe our methodology and the results.

\subsection{\boldmath Simulation tools and input parameters}
\label{sec:sim}

The simulation framework used to compute our results is as follows. We take the publicly-available UFO module \cite{wrufo} for the effective left-right-symmetric (LRS) model \cite{Atre:2009rg,Mattelaer:2016ynf,Mitra:2016kov}, which contains the heavy LRS gauge bosons $W_R^{\pm}$, $Z_R$ with masses $M_R$ and $M_{Z_R}$ and three heavy Majorana neutrinos $N_i$ with masses $M_i$, and we modify it by implementing non-universal couplings of the $W_R$ to third-generation quarks and interactions of $W_R^+W_R^-$ with the photon. We simulate parton-level events using \texttt{MadGraph5\_aMC@NLO} \cite{Alwall:2014hca}. These events are subsequently showered with \texttt{PYTHIA8} \cite{Sjostrand:2007gs,Sjostrand:2006za}. Jet clustering is then performed using \texttt{FastJet} \cite{Cacciari:2011ma}, implementing the anti-$k_t$ algorithm of Ref.~\cite{Cacciari:2008gp} with $R=0.4$. Detector simulation is perfomed using \texttt{Delphes3} \cite{deFavereau:2013fsa}.

\subsection{\boldmath Bounds from same-sign dilepton + jets searches at $\sqrt{s} = 7$ TeV}
\label{sec:SS2L}

Ref.~\cite{ATLAS:2012ak} describes the ATLAS search for hypothetical heavy neutrinos and RH gauge bosons in events with two reconstructed leptons and at least one hadronic jet. These experimental results were obtained from LHC data at $\sqrt{s} = 7$ TeV collected by the ATLAS detector, corresponding to an integrated luminosity of 2.1 fb${}^{-1}$. Here we recast these bounds to derive the 95\% C.L. exclusion regions in the $\lambda$-$M_R$ plane for various values of $M_1$. While this search is relatively old for LHC searches, it turns out to place the strongest constraints on the model.

The following parton-level NP processes contribute to the final state analyzed in the ATLAS paper: 
\begin{enumerate}
\item $pp\to eN_{1}\to eebj$,
\item $pp\to teN_{1}\to eebbjjj$,
\item $pp\to tbeN_{1}\to eebbbjjj$.
\end{enumerate}
In the limit $\lambda \ll 1$ the cross section of process 1 is proportional to $\lambda^2$, while processes 2 and 3 are independent of $\lambda$. In our simulations we fix $\lambda=0.042$.

Our analysis is performed at the particle level. For each configuration of the model parameters, we simulate 10k events using the simulation framework described in Sec.~\ref{sec:sim}. We then apply the ATLAS baseline selection to determine the ``acceptance $\times$ efficiency'' $({\cal A}\,\varepsilon )$ in the same-sign (SS) dilepton channel. The baseline selection consists of the following:
\begin{itemize}
\item only jets with $p_{T}>20$ GeV and $|\eta|<2.8$ are retained,
\item only leptons with $p_{T}>25$ GeV and $|\eta|<2.47$ are retained,
\item the closest jet to a lepton that has $\Delta R<0.5$ has been removed,
\item events with exactly two SS electrons and at least one jet are selected,
\item only events with $m_{ll}>110$ GeV are selected.
\end{itemize}
We compute the visible cross section of the three processes, i.e., cross sections $\times$ acceptance $\times$ efficiency ($\langle \sigma\,{\cal A}\,\varepsilon\rangle$),  for different values of $M_R$. In the small $\lambda$ limit, the total visible cross section as a function of $\lambda$ can be written as
\be 
\langle \sigma\,{\cal A}\,\varepsilon\rangle_{\rm tot}=\langle \sigma\,{\cal A}\,\varepsilon\rangle_1\left( \frac{\lambda}{0.042}\right)^2+\langle \sigma\,{\cal A}\,\varepsilon\rangle_2+\langle \sigma\,{\cal A}\,\varepsilon\rangle_3 ~.
\ee

\begin{figure}[h!]
\vspace{0.5cm}
\centering
\includegraphics[scale=0.5]{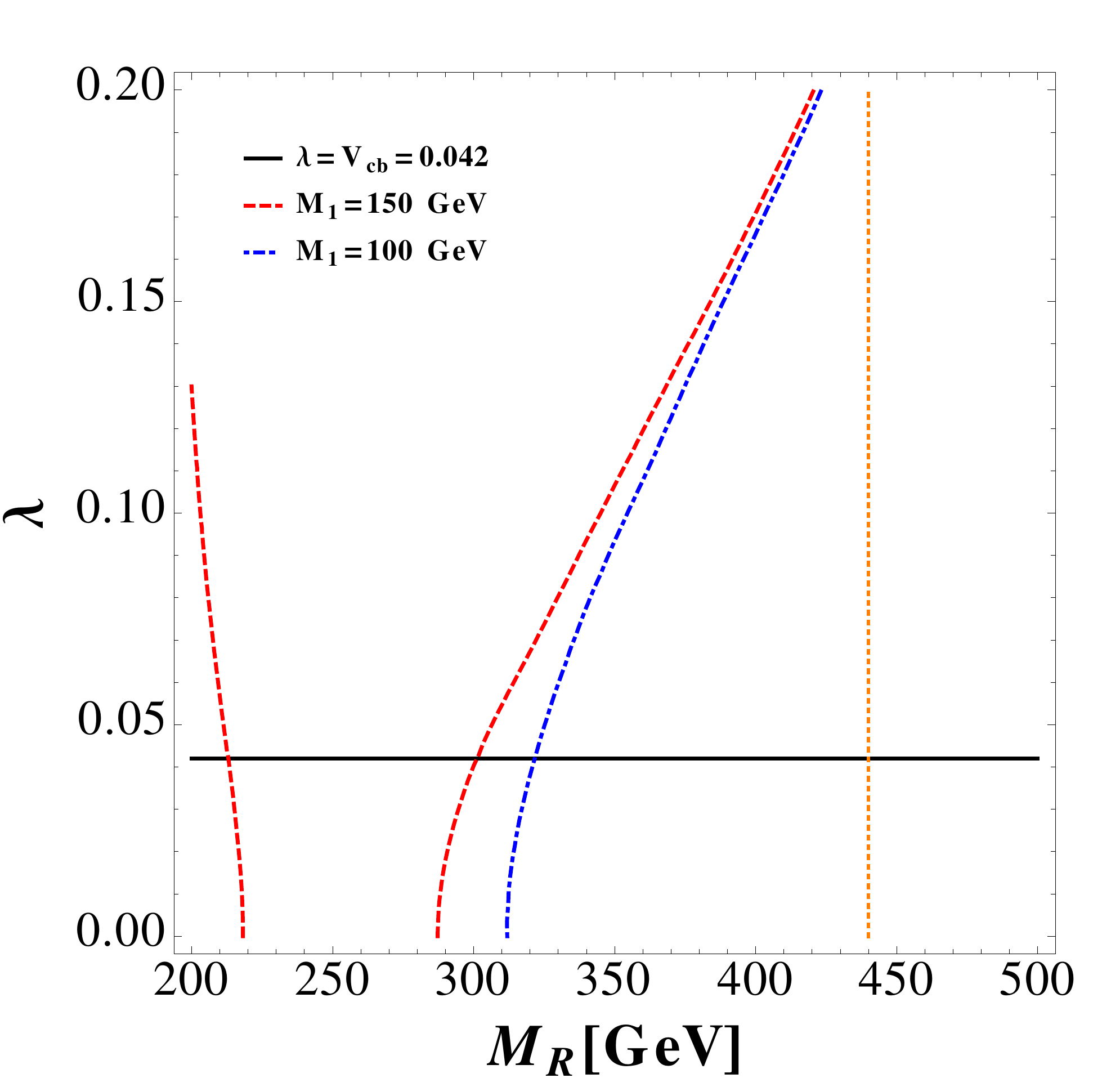} 
\caption{95\% C.L. bounds in the $\lambda$-$M_R$ plane obtained using the result of  the $e^{\pm}e^{\pm}$ channel in Table 3 of Ref.~\cite{ATLAS:2012ak}. For $M_1=150$ GeV, the excluded region is between the two dashed red 
lines. For $M_1=100$ GeV, the excluded region is to the left of the dot-dashed blue 
line. In this plot we also show the discovery region for $M_1=150$ GeV, which lies to the left of the vertical dotted orange 
line (see Sec.~\ref{sec:results}). The black horizontal line represents our choice of $\lambda=0.042$. 
\label{boundwr}}
\end{figure}

By inspection of Table 3 of the ATLAS paper \cite{ATLAS:2012ak}, we see that points in the $\lambda$-$M_R$ plane are excluded at 95\% C.L. if 
\be 
\langle \sigma\,{\cal A}\,\varepsilon\rangle_{\rm tot}>37.6\,\,{\rm fb}
\ee
The 95\% C.L. exclusion regions in the $\lambda$-$M_R$ plane for different values of $M_1$ are shown in Fig.~\ref{boundwr}. We see that for $M_1 \sim 100$ GeV and $\lambda =0$, this search excludes $M_R \lesssim 310$ GeV, and the bound gets moderately stronger with increasing $\lambda$. The situation is similar, though the bounds are slightly weaker, if $M_1$ is heavier, and there is also a region of lower $M_R$ that is allowed by this search when the electron in the decay of $W_R \rightarrow N e$ becomes too soft. For $M_1=50$ GeV the curve is similar to the case $M_1=100$ GeV. We were not able to get a curve for smaller masses because, in this case, the $N_1$ decay width becomes too small and the simulations turned out to be affected by numerical instabilities which do not allow us to get reliable results. 
In Fig.~\ref{boundwr} we also include a vertical line showing the discovery reach of our proposed search (Sec.~\ref{sec:results}), demonstrating that a substantial portion of the parameter space could yet be covered.

\subsection{\boldmath Bounds from multilepton final state searches at $\sqrt{s} = 13$ TeV}
\label{sec:multilepton}

A search for electroweak production of charginos and neutralinos in multilepton final states was performed by CMS \cite{CMS:2017moi}.  These experimental results were obtained from LHC data at $\sqrt{s} = 13$ TeV collected by the CMS detector, corresponding to an integrated luminosity of 35.9 fb${}^{-1}$. The bounds on our model are derived by considering  the search category G defined in Table 8 of Ref.~\cite{CMS:2017moi}, which is characterized by events with at least four light-flavour leptons, no hadronic taus, and the number of opposite-sign, same-flavour (nOSSF) leptons greater than or equal to two. Within this category, events are further divided into bins of $p_{T}^{\rm miss}$ (see Table 8 of Ref.~\cite{CMS:2017moi}). These results can be used to obtain 95\% C.L. exclusion limits on $M_R$.

The following parton-level NP process contributes to this search category:
\be 
pp\to e^{+}N_1e^{-}N_1\to e^{+}e^{+}e^{-}e^{-}b\,\bar b\,q\,\bar q ~,
\ee
where $q=u,c$. Our analysis is performed at the particle level. We fix $\lambda=0.042$, $M_1=100$ GeV and consider various values of $M_R$ between 200 and 350 GeV. For each value of $M_R$, we simulate 10k events using the simulation framework described in Sec.~\ref{sec:sim}. We then apply the CMS event selection to determine the expected signal yield in each $p_{T}^{\rm miss}$ bin. The selection consists of the following:
\begin{itemize}
\item only jets with $p_{T}>25$ GeV and $|\eta|<2.4$ are retained,
\item only isolated leptons such that $I_{mini}<0.4$ \cite{Rehermann:2010vq}, $p_{T}>10$ and $|\eta|<2.4$ are retained,
\item events with at least 4 isolated leptons are selected,
\item only jets with distance $\Delta R>0.4$ from each isolated lepton are retained,
\item events with at least 1 $b$-tagged jet are rejected,
\item events with nOSSF $\geq 2$ are selected.
\end{itemize}
At the particle level, in order to estimate the total number of events with no $b$-tagged jets, we consider the number of events $N_m$ with $m$ $b$-jets and multiply them by the probability of a real $b$-jet not being tagged which is taken to be $(0.37)^m$~\cite{CMS:2017moi}. The quantity $\sum_m N_m (0.37)^m$ represents our estimate of total events with no $b$-tagged jests. The number of background events $(b)$ and the number of observed events $(n^{obs})$ in each bin of $p_{T}^{\rm miss}$ is shown in Table 18 of Ref.~\cite{CMS:2017moi}, and repeated here in Table \ref{CMSpTbindata}.

\begin{table}[ht]
\centering
\begin{tabular}{|c|c|c|c|}
\hline 
 & $p_{T}^{miss}$ & $b$ & $n^{obs}$ \tabularnewline
\hline 
\hline 
1 & 0-50 & $b_{1}=460\pm120$ & $n_{1}^{obs}=619$ \tabularnewline
\hline 
2 & 50-100 & $b_{2}=45\pm14$ & $n_{2}^{obs}=51$ \tabularnewline
\hline 
3 & 100-150 & $b_{3}=2.7\pm0.8$ & $n_{3}^{obs}=2$ \tabularnewline
\hline 
4 & 150-200 & $b_{4}=1.12\pm0.33$ & $n_{4}^{obs}=2$ \tabularnewline
\hline 
5 & $>200$ & $b_{5}=0.97\pm0.32$ & $n_{5}^{obs}=0$ \tabularnewline
\hline 
\end{tabular}
\caption{The number of background events $(b)$ and the number of observed events $(n^{obs})$ in each $p_{T}^{\rm miss}$ bin (taken from Table 18 of Ref.~\cite{CMS:2017moi}).}
\label{CMSpTbindata}
\end{table}

We apply the CL$_{s}$ method~\cite{Read:2002hq} to extract bounds on $M_{R}$. To be specific, for each $p_{T}^{\rm miss}$ bin, we compute the signal+background distribution
as 
\begin{equation}
p(s_i+b_i,n_{i})=\frac{\int_{0}^{\infty}dze^{-\frac{(z-b_{i})^{2}}{2\Delta b_{i}^{2}}}e^{-(s_{i}+b_{i})}\frac{(s_{i}+b_{i})^{n_{i}}}{n_{i}!}}{\int_{0}^{\infty}dze^{-\frac{(z-b_{i})^{2}}{2\Delta b_{i}^{2}}}}
\end{equation}
and the background distribution as 
\begin{equation}
p(b_i,n_{i})=\frac{\int_{0}^{\infty}dze^{-\frac{(z-b_{i})^{2}}{2\Delta b_{i}^{2}}}e^{-b_{i}}\frac{b_{i}{}^{n_{i}}}{n_{i}!}}{\int_{0}^{\infty}dze^{-\frac{(z-b_{i})^{2}}{2\Delta b_{i}^{2}}}} ~.
\end{equation}
These definitions take into account the background uncertainty $\Delta b_i$ and assume its distribution to be Gaussian. Given the observed number of events in each bin, $n_{i}^{obs}$, we use the above distributions to compute the following $p$-values:
\begin{equation}
{\rm CL}_{s_i+b_i}=P(n_{i}<n_{i}^{obs}|s+b)=\int_0^{n_{i}^{obs}}dn_i\,p(s_i+b_i,n_{i}) ~,
\ee
\be 
{\rm CL}_{b_i}=P(n_{i}<n_{i}^{obs}|b)=\int_0^{n_{i}^{obs}}dn_i\,p(b_i,n_{i}) ~,
\end{equation}
and their ratio
\begin{equation}
{\rm CL}_{s_i}=\frac{{\rm CL}_{s_i+b_i}}{{\rm CL}_{b_i}} ~.
\end{equation}
Since the bin uncertainty correlations are unknown, we adopt a conservative approach and assign to each value of $M_{R}$ the smaller ${\rm CL}_{s}$ value among those computed for each bin, namely ${\rm CL}_{s}={\rm min}\{{\rm CL}_{s_i} \}$ . Values of $M_R$ are excluded at 95\% C.L. if 
\be 
{\rm CL}_{s} \leq 0.05 ~.
\ee

\begin{figure}[ht!]
\vspace{0.5cm}
\centering
\includegraphics[scale=0.3]{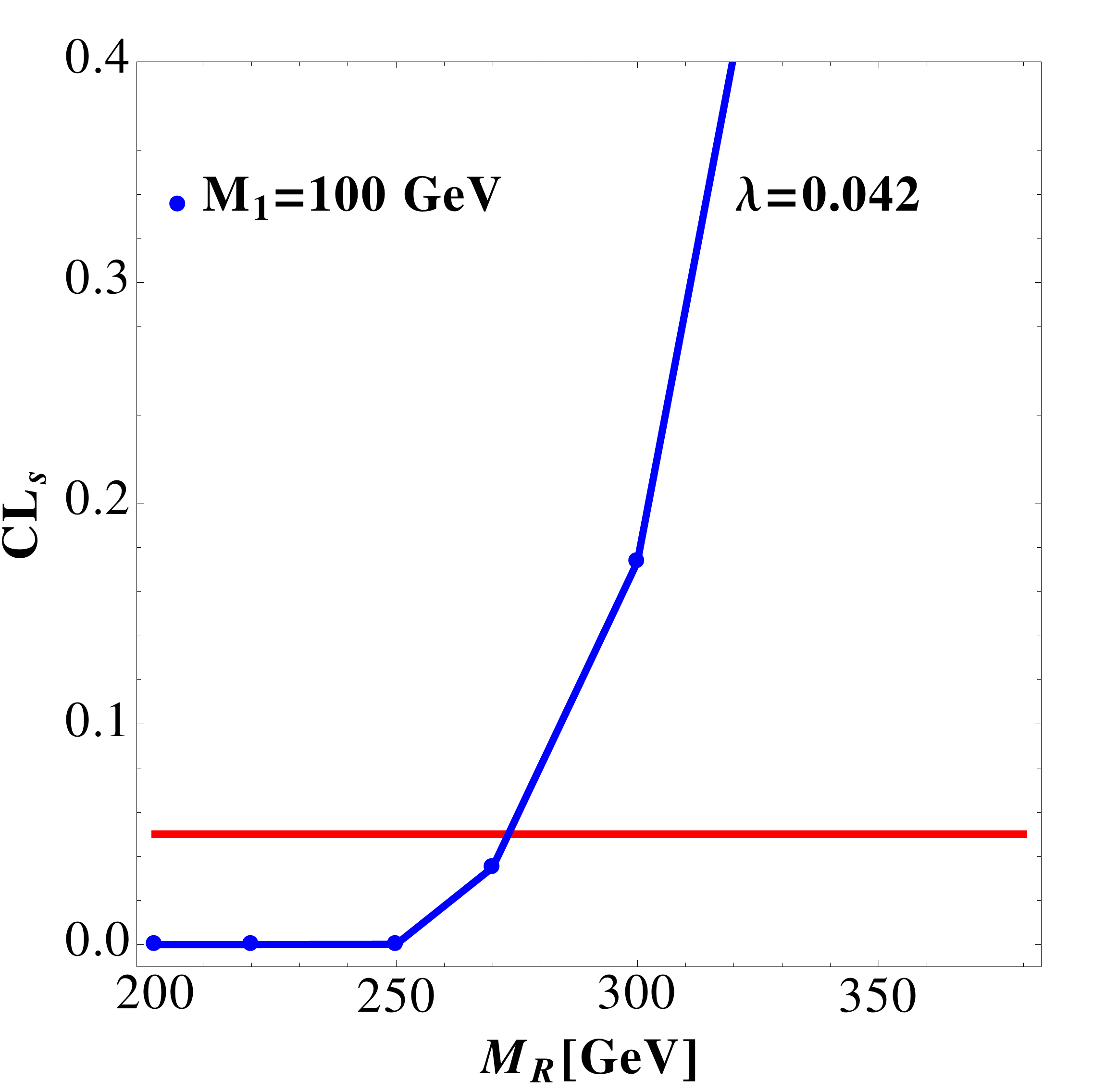} 
\includegraphics[scale=0.3]{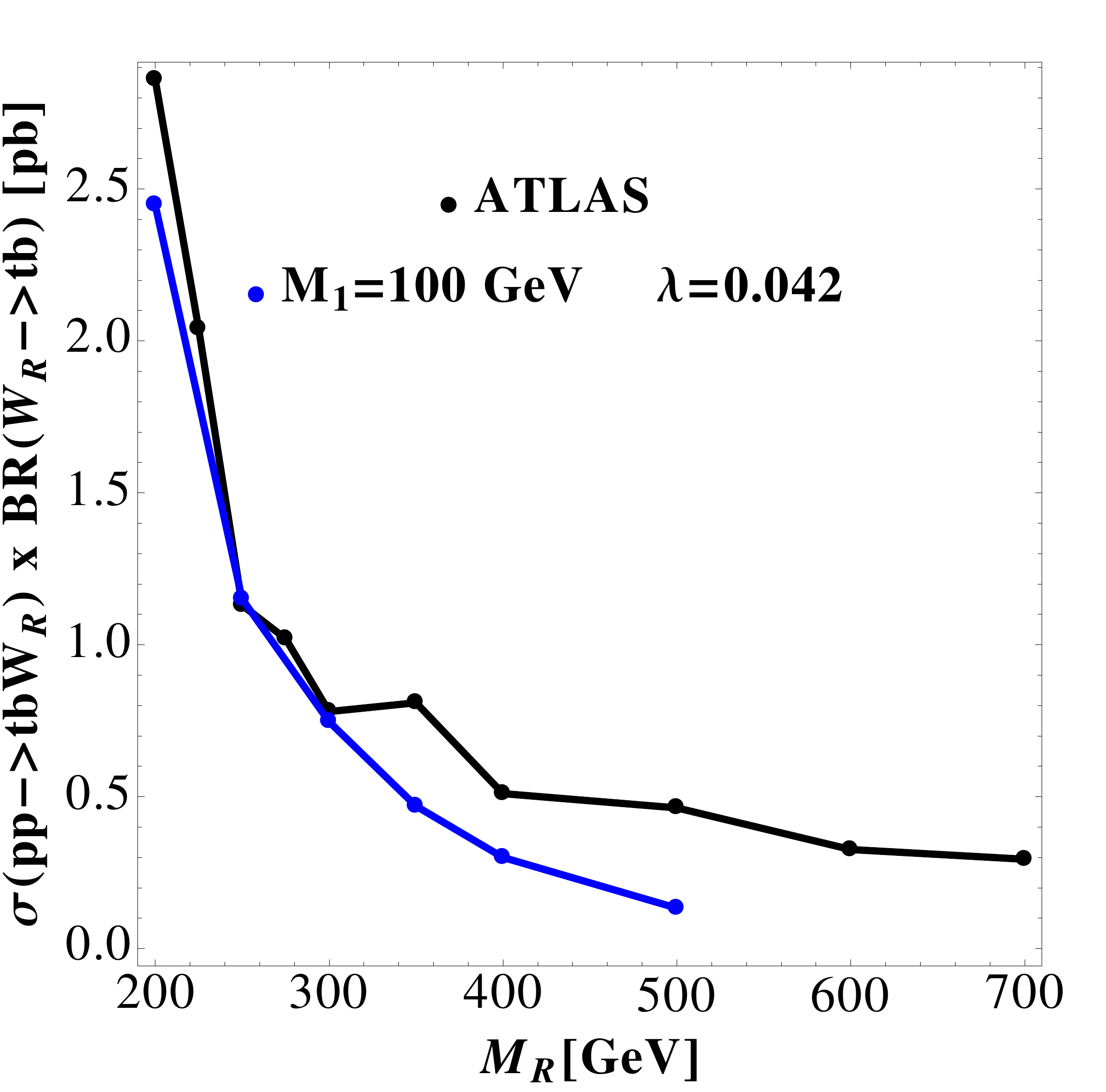} 
\caption{\textbf{Left:} ${\rm CL}_{s}$ exclusion limits on $M_R$ obtained by recasting the CMS multilepton search in Ref.~\cite{CMS:2017moi} using the G category. Masses for which ${\rm CL}_{s}<0.05$ are excluded at 95\% C.L. \textbf{Right:} 95\% C.L. exclusion limits of $M_R$, obtained by recasting the ATLAS search for heavy charged Higgs bosons $H^\pm$ decaying via $H^\pm\to tb$ \cite{ATLAS:2018ntn}. All bounds are obtained for $\lambda=0.042$ and $M_1=100$ GeV.} 
\label{clsi}
\end{figure}

In Fig.~\ref{clsi}, the left panel shows the ${\rm CL}_{s}$ values as a function of $M_{R}$. We find that values of $M_R \lesssim 280$ GeV are excluded at 95\% CL. We have computed this bound just for $M_1=100$ GeV since it is weaker than the bound imposed by same-sign dilepton + jet searches at $\sqrt{s} = 7$ TeV presented in the previous subsection.

\subsection{\boldmath Bounds from associated charged-resonance production searches at $\sqrt{s} = 13$ TeV}
\label{sec:tb}

The ATLAS search for heavy charged Higgs bosons $H^\pm$ decaying via $H^\pm\to tb$ is described in Ref.~\cite{ATLAS:2018ntn}. These experimental results were obtained from LHC data at $\sqrt{s} = 13$ TeV collected by the ATLAS detector, corresponding to an integrated luminosity of 36.1 fb${}^{-1}$. They can be used to derive 95\% C.L. exclusion limits on $M_R$, assuming the spin of the decaying resonance does not significantly change the acceptance of the search. This should be a reasonable approximation since this is a 2-body decay.

Figure 8 of Ref.~\cite{ATLAS:2018ntn} shows the observed 95\% C.L. limits for the production cross section $\sigma(pp\to tbH^{\pm})\times BR(H^{\pm}\to tb)$ as a function of $M_H$. We fix $\lambda=0.042$, $M_1=100$ GeV and use  \texttt{MadGraph5\_aMC@NLO} to compute the cross section $\sigma(pp\to tbW_{R}^{-})\times BR(W_{R}^{-}\to tb)$ for different values of $M_R$. We then take the experimental results in Fig. 8 of Ref.~\cite{ATLAS:2018ntn} to set some constraints on $M_{R}$. The results are shown in the right panel of Fig.~\ref{clsi}. From this plot we see that there is no bound coming from this search. 

\subsection{\boldmath Bounds from resonances decaying to top and bottom quarks at $\sqrt{s} = 1.96$ TeV}
\label{sec:tbTev}

In Ref.~\cite{CDF:2015aej}, the CDF II detector placed bounds on charged massive resonances decaying to top and bottom quarks ($p \bar{p} \to W' \to t b$) . These results use the full data set at the Tevatron ($\sqrt{s} = 1.96$ TeV), with an integrated luminosity of $9.5\, \text{fb}^{-1}$, and can be used to derive 95\% C.L. constraints on $M_R$.

\begin{figure}[ht!]
\vspace{0.5cm}
\centering
\includegraphics[scale=0.4]{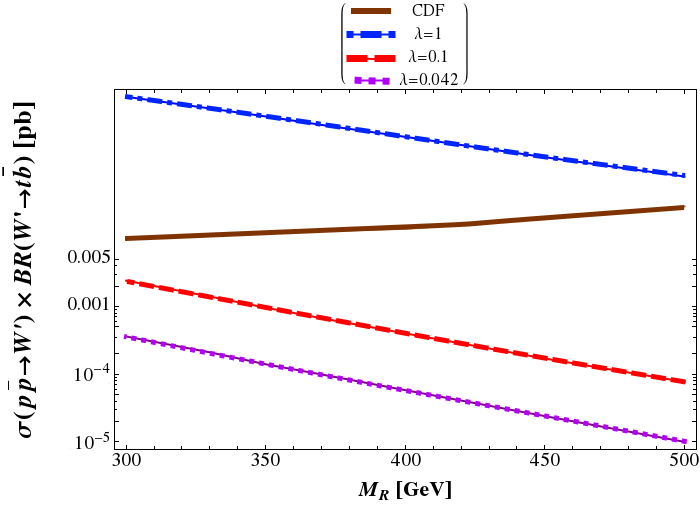} 
\caption{Bound on $\sigma (p \bar{p}\to W_{R}^{+})\times BR(W_{R}^{+}\to t \bar{b})$ as a function of $M_R$ for $\lambda=1$ (blue, dash-dotted line), $\lambda=0.1$ (red, dashed line) and $\lambda=0.042$ (purple, dotted line). The solid brown curve represents the upper experimental limit obtained by the CDF II detector data \cite{CDF:2015aej}, with an integrated luminosity of $9.5\, \text{fb}^{-1}$ and a center-of-mass energy of 1.96 TeV.} 
\label{CDFbounds}
\end{figure}

We use  \texttt{MadGraph5\_aMC@NLO} to compute the quantity $\sigma(p \bar{p}\to W_R^+)\times BR(W_R^+\to t \bar{b})$. As noted above, for resonant production the cross section will scale like $\lambda^2$.  The experimental results of Ref.~\cite{CDF:2015aej} start at $M_R = 300$ GeV, and we show the cross section for three different benchmarks for $\lambda$: 1, 0.1 and 0.042. and compare them to the limit given in Fig.~2 of Ref.~\cite{CDF:2015aej}. The upper limit on $\sigma (p \bar{p}\to W_R^+)\times BR(W_R^+\to t \bar{b})$ as a function of $M_R$ in Fig.~\ref{CDFbounds}, from which we conclude that $\lambda=1$ is excluded over the entire mass region. Using the fact that $\sigma \propto \lambda^2$, we find that there are no bounds on $M_R$ above 300 GeV for $\lambda \lesssim 0.18$. For this process, the value of the right handed neutrino mass $M_1$ only matters for computing the branching ratio of the $W_R \rightarrow tb$, and the bounds will approximately independent of $M_1$ as long $M_1$ is sufficiently below than $M_R$. When the two masses are approximately degenerate, the bounds shown in Fig.~\ref{CDFbounds} will get somewhat stronger.

\section{Discovery potential}
\label{sec:results}

Above, we described several processes that lead to two same-sign electrons in the final state. What is particularly intriguing is that one of them can lead to three same-sign electrons (3SSe) + jets. It is the LNV3 process
\bea
&& pp\to e^{+}N_1e^{-}N_1\to e^{+}e^{-}e^{+}e^{+}b\,b\,\bar q\,\bar q ~, \nn\\
&& pp\to e^{+}N_1e^{-}N_1\to e^{+}e^{-}e^-e^-\bar b\,\bar b\, q\, q ~.
\eea
where $q=u,c$. This is a striking signal of NP; here we estimate the discovery potential of our model at the LHC13 using this mode.
Note that, although we are focusing on the 3SSe signal as a way of detecting our simplified $W_R$ model, it is more general than this. This same LNV signal can be generated in other NP models (for example, see Refs.~\cite{Ozcan:2009qm, Mukhopadhyaya:2010qf, Bambhaniya:2013yca, Guo:2017ybk, Agarwalla:2018xpc}).

\subsection{\boldmath SM backgrounds}
\label{sec:bckg}

We begin by identifying the SM processes that constitute the main backgrounds for the 3SSe signal events. These are $p p \to t\, \bar t$, $p p \to t\, \bar t\, t\, \bar t$, $p p \to t\, \bar t\, l^+\, l^-$, $p p \to t\, \bar t\, W^\pm$, $p p \to t\, \bar t\, V\, V$, $p p \to W^\pm\, W^\pm\,  j\, j$, $p p \to V\, W^\pm$, $p p \to  V\, V\, V$, $p p \to  V\, V\, V\, V$ and $p p \to l^+\, l^-\, l^+\, l^-$, where $V=Z,W^+,W^-$ and are summarized in Tab.~\ref{BGtab}. Some of these processes have 3SSe events directly at parton level, while others require additional fake electrons coming from some hadronic activity and/or charge mismeasurement effects at reconstruction level in order to contribute to the 3SSe final state. We will use our detector simulation to estimate the fake rate of leptons, but for a theoretical estimate see~\cite{Curtin:2013zua}.

To determine the size of each background, we use \texttt{MadGraph5\_aMC@NLO} to simulate $N_\text{ev}$ events of the SM process $p p \to f $ (where $f$ is one of the final states mentioned above) at $\sqrt{s} = 13$ TeV. We then shower the events with \texttt{PYTHIA8} and run \texttt{Delphes3} in order to simulate detector effects, using the default parameter card. The cross section for a specific SM decay to have 3SSe can be written as 
\be 
\label{BG}
\sigma (\text{3SSe})_f \approx \sigma_{\text{SM}} (p p \to f)\times
\frac{N_\text{3SSe}+\varepsilon_c\times (N_\text{3e}+4\times N_{2e^+2e^-}) }{N_\text{ev}} ~,
\ee
where $\sigma_{\text{SM}} (p p \to f)$ is the SM production cross section of the final state $f$, $N_\text{3SSe}$ is the number of events obtained with 3SSe and $N_\text{ev}$ is the total number of events generated. The $N_{\text{3SSe}}$ term includes both real and fake electrons. The $N_\text{3e}$ and $N_{2e^+2e^-}$ terms parameterize the effects of charge mismeasurement.  In general charge mismeasurement (CM) is a complicated function of $p_T$ and $\eta$ and also can be tuned by adjusting lepton acceptance probability~\cite{CMS:2020uim,CMS:2019ied,ATLAS:2019jvq}. We will take an average charge mismeasurement probability of $\varepsilon_c \approx 10^{-3}$ which is consistent with the experimental measurements of this parameter, particularly those at CMS~\cite{CMS:2020uim,CMS:2019ied}. 
$N_\text{3e}$ is the number of simulated events with three electrons not all of the same sign (i.e., $2e^+e^- + 2e^-e^+$),  and $N_{2e^+2e^-}$ is the number  of events with two $e^+e^-$ pairs. 

For processes with very large production cross sections, we also require that the heavy SM states decay to electrons to make the Monte Carlo statistics more manageable. For $t\, \bar t$, we require the decay as $t\, \bar t\to b e^+ \nu_e \bar{b} e^- \bar{\nu}_e$, $b e^+ \nu_e \bar b jj$, or $\bar{b} e^- \bar{\nu}_e b jj$, while for $f=W^\pm\, W^\pm\,  j\, j$, $W^+\, W^-$ and $Z\, W^\pm$, we require $W^\pm \to e^\pm \overset{\scriptscriptstyle(-)}{\nu_e}$ and $Z\to e^+e^-$. In Eq.~\eqref{BG}, $\sigma_{\text{SM}} (p p \to f)$ must then be multiplied by the appropriate branching ratios.

For each of the SM background processes, all the relevant simulation parameters are summarized in in Table~\ref{BGtab}. The process cross sections and their cross section for generating 3SSe are then given in Table~\ref{background}. For most processes, the background cross section is dominated by charge mismeasurement, so we also give the right column of Table~\ref{background} of the rate if CM is set to zero. This shows that if experiments are able to tune their lepton selection procedures to reduce CM, the background rate in this search would go down significantly.

\begin{table}
\centering
\begin{tabular}{|c|c|c|c|c|c|}
\hline 
 Background process & Decay channels considered & $N_\text{ev}$ & $N_\text{3SSe}$  &  $N_\text{3e}$ & $N_{2e^+2e^-}$  \tabularnewline
\hline 
\hline 
$p p \to t\, \bar t$ & $t \to b e^+ \nu_e, \, \bar{t} \to \bar{b} e^- \bar{\nu}_e$ & $10^5$ & 0  &  13 & 5  \tabularnewline
\hline 
$p p \to t\, \bar t\, t\, \bar t$ & Inclusive & $5\times 10^5$ & 2  & 311  & 3  \tabularnewline
\hline
$p p \to t\, \bar t\, l^+\, l^-$ & Inclusive  & $2\times 10^5$  & 4  & 4318 & 148 \tabularnewline
\hline
$p p \to t\, \bar t\, W$ & Inclusive & $10^6$  & 1  & 276 & 1  \tabularnewline
\hline 
$p p \to t\, \bar t\, V\, V$ & Inclusive & $3\times 10^5$  &  5 & 202 & 3  \tabularnewline
\hline 
$p p \to W^\pm\, W^\pm\,  j\, j$ & $W^\pm \to e^\pm \overset{\scriptscriptstyle(-)}{\nu_e}$& $5\times 10^5$  & 35  & 42 & 0  \tabularnewline
\hline
$p p \to V\, W^\pm$ & $W^\pm \to e^\pm \overset{\scriptscriptstyle(-)}{\nu_e}, \, Z \to e^+ e^-$ & $3\times 10^5$  &  3 & 4582 & 5  \tabularnewline
\hline 
$p p \to  V\, V\, V$ & Inclusive & $5\times 10^5$  & 3  & 436 & 23 \tabularnewline
\hline 
$p p \to  V\, V\, V\, V$ & Inclusive & $10^5$  & 8  & 248 & 13  \tabularnewline
\hline
$p p \to l^+\, l^-\, l^+\, l^-$ & Inclusive & $3\times 10^5$  & 0 & 30482 & 20504 \tabularnewline
\hline
\end{tabular}
\caption{For each SM background process of the first column, we generate $N_\text{ev}$ events. We obtain $N_\text{3SSe}$ events with final states containing three same-sign electrons, $N_\text{3e}$ events with final states containing three electrons, not all of the same sign, and $N_{2e^+2e^-}$ events with final states containing two $e^+e^-$ pairs. The second column indicates whether the produced state $f$ of the first column decays to a specific  final state or to all possible final states (inclusive).}
\label{BGtab}
\end{table}

\begin{table}[ht]
\centering
\begin{tabular}{|c|c|c|c|}
\hline 
Background process & $\sigma_{\text{SM}} (p p \to f) [fb]$ & {$\sigma(\text{3SSe})_f$ [fb]}  &  { $\sigma(\text{3SSe})_f$ (no CM) [fb]} \tabularnewline
\hline 
\hline 
$p p \to t\, \bar t$& 9780  \cite{Catani:2019hip,ParticleDataGroup:2020ssz} & $\sim 5\times 10^{-3}$ &    $\sim 2\times 10^{-3}$\tabularnewline
\hline 
$p p \to t\, \bar t\, t\, \bar t$& 12  \cite{CMS:2019rvj,ATLAS:2020hpj} & $6\times 10^{-5}$ &    $5\times 10^{-5}$\tabularnewline
\hline
$p p \to t\, \bar t\, l^+\, l^-$& 60.8  \cite{ttllBG}  & $3\times 10^{-3}$ &    $1\times 10^{-3}$\tabularnewline
\hline
$p p \to t\, \bar t\, W$ &  722.5  \cite{Frederix:2021agh}& $1\times 10^{-3}$ &   $0.7\times 10^{-3}$\tabularnewline
\hline 
$p p \to t\, \bar t\, V\, V$& 18.114  \cite{Maltoni:2015ena}& $3\times 10^{-4}$ &   $3\times 10^{-4}$\tabularnewline
\hline 
$p p \to W^\pm\, W^\pm\,  j\, j$ & 0.05  \cite{CMS:2017fhs,ParticleDataGroup:2020ssz} & $4\times 10^{-6}$ &    $ 4\times 10^{-6}$\tabularnewline
\hline
$p p \to V\, W^\pm$& 1649.5  \cite{ATLAS:2017bbg,ATLAS:2019bsc,ParticleDataGroup:2020ssz} & $4\times 10^{-2}$ &    $ 2\times 10^{-2}$\tabularnewline
\hline 
$p p \to  V\, V\, V$ & 991.7  \cite{CMS:2020hjs} & $7\times 10^{-3}$ &    $6\times 10^{-3}$\tabularnewline
\hline 
$p p \to  V\, V\, V\, V$ & 2.2  \cite{Hussein:2006er} & $2\times 10^{-4}$ &    $2\times 10^{-4}$\tabularnewline
\hline
$p p \to l^+\, l^-\, l^+\, l^-$& $56.49$  \cite{Grazzini:2021iae}  & $2\times 10^{-2}$ &   $\lesssim 2 \times 10^{-4}$\tabularnewline
\hline\hline
Total & & $8\times 10^{-2}$&  $3\times 10^{-2}$\tabularnewline
\hline
\end{tabular}
\caption{SM background cross sections, with and without charge mismeasurement, computed for $\sqrt{s} = 13$ TeV.  In our calculations, we use the central values of $\sigma_{\text{SM}} (p p \to f)$ and take into account the branching ratios of the decay modes presented in the second column of Table~\ref{BGtab}. 
}
\label{background}
\end{table}

We notice from Table~\ref{BGtab} that $N_\text{3SSe}=0$ in the first and last lines, which means that the Monte Carlo statistics were not sufficient. Because the cross sections for those two processes are very large, we need a more precise estimate of $N_\text{3SSe}$.
For $p p \to t\, \bar t$, we need to estimate the background without charge mismeasurement. From the experimental paper~\cite{CMS:2020cpy} searching for same-sign dileptons, we find that there are $\sim$ 5000 events with same-sign dileptons from non-prompt or fake leptons. If we assume that they are dominated by semi-leptonic $t\, \bar{t}$ decays, then the rate of getting one fake lepton is $\sim  1.3\times 10^{-4}$. We then square that fake rate since we need 2 fake electrons to obtain a 3SSe signal,\footnote{To be conservative, we assume that all fake leptons are electrons or positrons.} and multiply this rate by the cross-section $\sigma (p p \to t\, \bar t\to b e^+ \nu_e \bar b jj, \, \bar{b} e^- \bar{\nu}_e b jj)$. We obtain the rate of detecting 3SSe without charge mismeasurement of order $\sim 2 \times 10^{-3}$ fb. Therefore the total rate of the $t\, \bar{t}$  background is $\sim 5 \times 10^{-3}$ fb.  
For the process $p p \to  l^+l^-l^+l^-$, we can estimate the cross-section for having 3SSe without charge mismeasurement by assuming that we get at the most one 3SSe event, so that this cross-section does not exceed $\sigma \times N_{\text{ev}} \sim 2 \times 10^{-4} \text{ fb}$.


\subsection{NP signal}

\begin{figure}[ht!]
\vspace{0.5cm}
\centering
\includegraphics[scale=0.4]{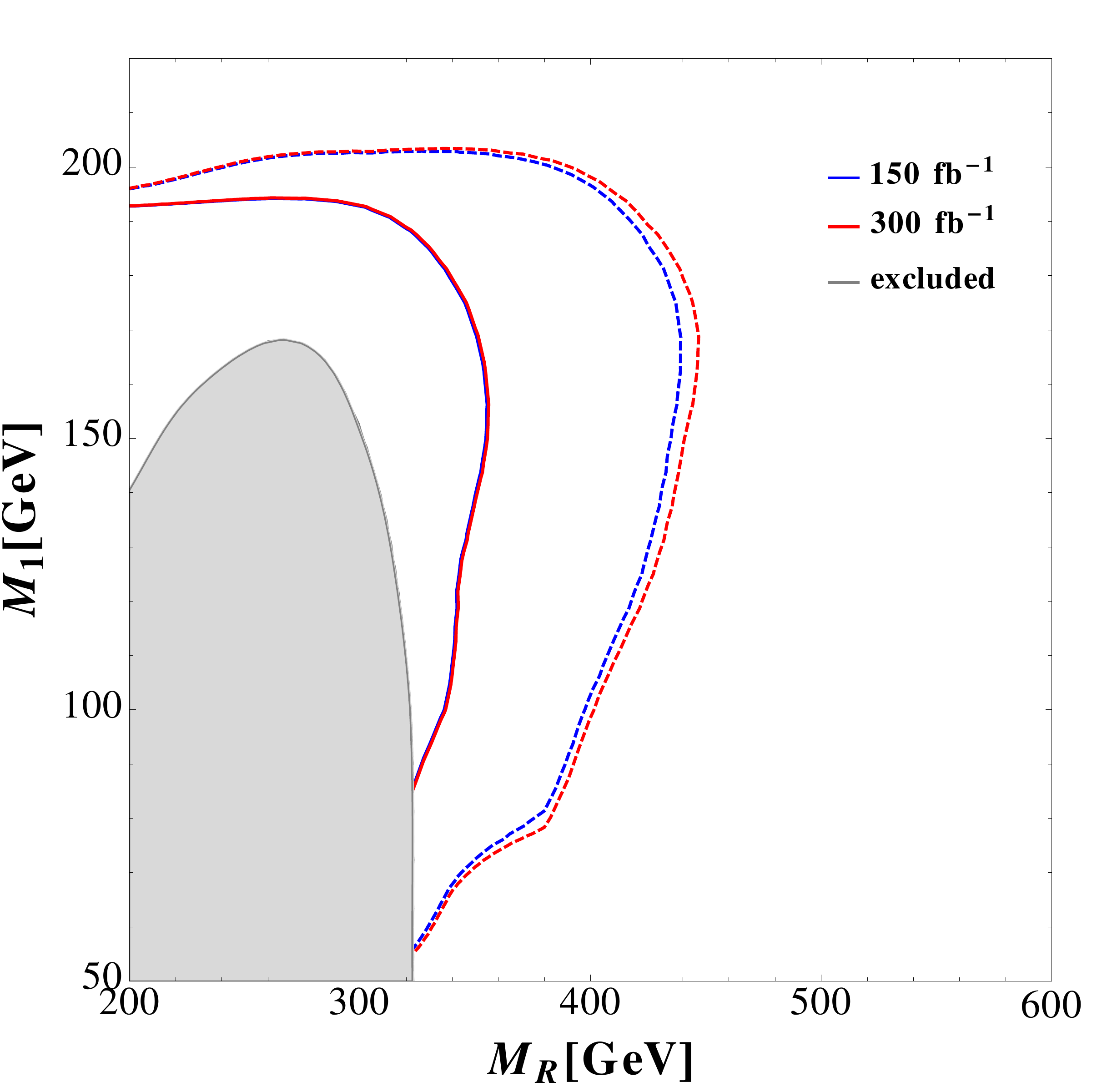} 
\caption{NP discovery region for the LHC at $\sqrt{s} = 13$ TeV. The blue (red) curve represents the results for 150 (300) fb${}^{-1}$. Points that lie to the left of the blue and red curves have an expected discovery significance of $\ge 5\sigma$. The solid curves consider backgrounds from fake electrons and charge mismeasurements, while the dashed curves consider only backgrounds from fake electrons and no charge mismeasurements. The gray region is excluded by the searches described in section~\ref{sec:constraints}. Here $\lambda = 0.042$.}
\label{sensitivity}
\end{figure}

In order to estimate the expected number of signal events, we perform an analysis at the reconstruction level, taking into account detector effects. With  \texttt{MadGraph5\_aMC@NLO}, we simulate 10k events for different values of $M_R$ and $M_1$, and we subsequently shower them with \texttt{PYTHIA8}. The detector simulation is performed by running \texttt{Delphes3} with the default parameter card, as was done for the background.  In the simulations, we fix $\lambda=0.042$ (in the limit of $\lambda\ll 1$ the cross section is independent of $\lambda$), and for each point in the $M_1$-$M_R$ plane we determine the visible cross section for producing 3SSe in the final state at the reconstruction level. We then compute the expected number of signal events ($S$) for two different luminosities, 150 fb${}^{-1}$ and 300 fb${}^{-1}$. As was shown in the previous subsection, the SM background events ($B$) are mainly due to fake leptons and electron charge mismeasurements. In order to discover the 3SSe signal at the $5\sigma$ level, we require
\be 
\frac{S}{\sqrt{B+\sigma_B^2}}\geq 5 ~,
\label{eq:disc}
\ee
where $\sigma_B$ represents the background uncertainty. We conservatively take this uncertainty to be 100\%, i.e., $\sigma_B=B$. The results are shown in Fig.~\ref{sensitivity}, which shows discovery sensitivity in the $M_1$-$M_R$ plane. 
The blue and red curves represent the results for 150 and 300 fb$^{-1}$, respectively; the points where the expected discovery significance is $\ge 5\sigma$ lie to the left of these curves. An exclusion at $2\sigma$ will be possible over a larger region of parameter space. We see that a significant region of parameter space not excluded by the searches recast in section~\ref{sec:constraints} could be discovered with the method proposed here. 
The expected discovery reach only has a very limited increase with increased luminosity. This is because the significance parameter in Eq.~\eqref{eq:disc} is dominated by the systematic uncertainty on the background, which we have conservatively estimated to be 100\% of the background. If this uncertainty can be reduced by data-driven or other methods, the scaling with increased luminosity will become more favorable. 

For our background analysis, we estimated the charge mismeasurement probability to be $\varepsilon_c\sim 10^{-3}$. This is probably an upper limit; in the actual experiment, this probability may well be smaller. For this reason, in Fig.~\ref{sensitivity} we present two cases: the solid blue and red curves consider backgrounds from both fake electrons and charge mismeasurements, while the dashed curves consider only backgrounds from fake electrons and no charge mismeasurements. The difference between the two regions to the left of these curves, which is probably where the true discovery region lies, is sizeable.

We emphasize once again that our simplified $W_R$ model is presented as an example of light NP particles that could have evaded detection. As we see here, the presence of such particles {\it can} be detected if one looks in the right place.

\section{Conclusions}
\label{sec:conc}

Despite the best efforts of experimentalists around the globe, no physics beyond the SM has been discovered at the LHC. In light of this, new-physics effects are now typically parameterized using an effective field theory, which is applicable at the LHC only if the mass scale of new physics is several TeV or larger. Even so, it is still possible that there are light new particles, but with properties that enable them to have escaped detection. It is incumbent upon theorists to construct models of such new physics, and to find ways of observing it.

In this paper, we consider a simplified model containing a right-handed $W$ boson in which the $W_R^+$ couples predominantly to $t{\bar b}$ in the quark sector. The $W_R^+$ also couples to $c{\bar b}$, but this coupling is assumed to be suppressed by a factor $\lambda \sim |V_{cb}| \approx 0.042$ compared to the $t\bar b$ coupling; all other quark couplings are even more suppressed. In the lepton sector, the model includes three right-handed neutrinos, which are assumed to have hierarchical masses $M_1\ll M_2,M_3$, with $M_1\sim {\cal O}(100)$ GeV. For simplicity, we assume that the $N_1$ is coupled to a single flavour of charged leptons, which we take to be the electron.

Because the $W_R$ does not couple to light quarks, it is mainly produced in association with heavy quarks, or pair-produced via a photon. It decays to $tb$ and to $e N_1$, with the $N_1$ decaying further to $ecb$. Because the neutrino is Majorana, it decays equally to $e^- c {\bar b}$ and $e^+ {\bar c} b$. This means that, half the time, the final state will contain two same-sign electrons, a signal of new physics. 

A thorough search of the literature revealed a number of recent LHC measurements whose results could be translated into constraints on the $W_R$. As an example, experiments have looked for two same-sign final-state leptons as a signal of heavy Majorana neutrinos. Taking all bounds into account, we find that the $W_R$ can still be as light as $\sim 300$ GeV.

In the $W_R$ pair-production channel, if each $W_R$ decays to $e N_1$, because the $N_1$ is a Majorana particle, one can have a final state with three same-sign electrons. This distinct lepton-number-violating final state is the signal of our model. We carefully estimated the SM background for this channel, and we find that the signal can be seen at the LHC in a significant region of parameter space, even with conservative assumptions about the background. We believe that a dedicated experimental search could have even better reach.  

While we have assumed that the lightest $N_1$ couples to the electron, there are of course other possibilities. If the $N_1$ couples to $\tau$, then the signal becomes much more difficult to detect experimentally. On the other hand, if it couples to muons, the background is significantly reduced, as both the fake and charge mismeasurement rates for muons are significantly smaller than for electrons. While we have not done a complete study, we expect the muon case would be nearly background free. Finally, if $N_1$ couples to multiple flavours, the parameter space is more complicated to analyze, but we have exploited the fact that the SM is flavour democratic while the signal couples to a single flavour. Therefore, if the $N_1$ couples to multiple flavours, the discovery reach is likely weakened.

We must stress that our simplified $W_R$ model is presented as an example of new physics that could have evaded detection and can therefore still be light. As we see here, it is still possible to detect the presence of such new particles if one looks in the right place.

\section*{Acknowledgments}
This work is supported in part by the Natural Sciences and Engineering Research Council of Canada (NSERC) and the Arthur B. McDonald Canadian Astroparticle Physics Research Institute.

\appendix
\section{\boldmath EFT limit}
\label{sec:EFT}

The language of the Standard Model Effective Field Theory (SMEFT) is very suitable for phenomenological studies in the presence of heavy new physics. The lack of discovery of new physics at the LHC has led to significant activity in parameterizing new-physics searches at the LHC in terms of EFTs. In this appendix, we explore how the model analyzed in this work connects to an effective field theory (EFT), and further describe how such an EFT is not suitable for analysis at the LHC. 

For $M_R \sim 300$ GeV, we can consider the $W_R$ a heavy field and parameterize its affects in terms of higher-dimensional operators. On the other hand, the $N_1$ is lighter than, for example, the top quark, so it is kept in the theory. Of course, this is not a clear seperation of scales, and so will contribute to the breakdown of the EFT. The EFT of the SM fields + a right-handed neutrino is typically called $\nu$SMEFT and has been explored in the literature~\cite{delAguila:2008ir,Aparici:2009fh,Bhattacharya:2015vja,Liao:2016qyd}.

There are three classes of dimension-six operators generated by integrating out the $W_R$:
\begin{eqnarray}
&&\frac{c_{duNe}}{\Lambda^{2}}\left(\bar{d}_{R}\gamma_{\mu}u_{R}\right)\left(\bar{N}_{R}\gamma^{\mu}e_{R}\right)+{\rm h.c.,} \nn\\
&&\frac{c_{ud}^{(1)prst}}{\Lambda^{2}}\left(\bar{u}_{Rp}\gamma_{\mu}u_{Rr}\right)\left(\bar{d}_{Rs}\gamma^{\mu}d_{Rt}\right)+{\rm h.c.,}
\nn\\
&&\frac{c_{eN}}{\Lambda^{2}}\left(\bar{e}_{R}\gamma_{\mu}e_{R}\right)\left(\bar{N}_{R}\gamma^{\mu}N_{R}\right) ~.
\label{eq:eftwr}
\end{eqnarray}
We can perform a tree-level matching computation to get the LO Wilson coefficients including the full flavour structure. The coefficient of the purely leptonic operator is given by
\begin{equation}
   \frac{c_{eN}}{\Lambda^{2}}=-\frac{ (k_{l})^2g^{2}}{2M_{W_{R}}^{2}} \, , 
\end{equation}
while the coefficients of those operators involving only third-generation quarks are given by
\begin{equation}
\frac{c_{btNe}}{\Lambda^{2}}=-\frac{ k_{q}k_{l}g^{2}}{2M_{R}^{2}}\qquad \qquad
\frac{c_{ud}^{(1)ttbb}}{\Lambda^{2}}=-\frac{ (k_{q})^2g^{2}}{2M_{R}^{2}}
\, .
\label{eq:wilson-big}
\end{equation}
Effective-field-theory analyses typically assume $\Lambda \sim$ TeV and $c_i\lesssim {\cal O}(1)$. This cannot be the case for our model because the $M_R$ can be as light as $\sim 300$ GeV, which means that if one were to choose $\Lambda =$ 1 TeV, this would imply very large values for the dimensionless Wilson coefficients $c_i$ in Eq.~\eqref{eq:wilson-big}. In other words, those points would lie outside the validity of the EFT. 

Operators involving one second-generation quark have coefficients suppressed by $\lambda$ compared to those in  Eq.~\eqref{eq:wilson-big}:
\begin{equation}
    \frac{c_{bcNe}}{\Lambda^{2}}=\lambda \frac{c_{btNe}}{\Lambda^{2}}
    \qquad
    \frac{c_{ud}^{(1)tcbb}}{\Lambda^{2}}=\lambda\frac{c_{ud}^{(1)ttbb}}{\Lambda^{2}}
    \label{eq:wilson-small}
\end{equation}
The natural scale for these operators is then $\Lambda \sim M_R / \sqrt{\lambda} \gtrsim $ TeV if we use $\lambda \approx V_{cb}$. The operators involving one first-generation quark or two second-generation quarks will be suppressed by $\lambda^2$ relative to those in Eq.~\eqref{eq:wilson-big}. Therefore, in this model, a naive guess of $\Lambda \gtrsim$ TeV will break down for $c_{eN}$, $c_{btNe}$ and $c_{ud}^{(1)ttbb}$, but will be sufficient for the remaining operators. 

In order to determine the breakdown of the EFT, we consider $M_R=200$ GeV, $M_1=100$ GeV, $k_l=k_q=1$ and $\lambda=0.042$. In our simplified model, this point is excluded by the ATLAS same-sign dilepton + jets searches at $\sqrt{s} = 7$ TeV, as shown in Fig.~\ref{sensitivity}. If we use the EFT matching conditions in Eqs.~\eqref{eq:wilson-big} and~\eqref{eq:wilson-small} to estimate the size of the Wilson coefficients $c_{btNe}/\Lambda^{2}$ and $c_{bcNe}/\Lambda^{2}$, and compute the cross sections $pp\to eN_{1}$, $pp\to teN_{1}$, $pp\to tbeN_{1}$ using the EFT model, we obtain values that are roughly two orders of magnitude smaller than the cross sections computed in the simplified model. This means that no limits can be placed on the EFT coefficients. This signals a breakdown of the EFT limit, as expected from the fact that the cutoff of this theory is $\sim M_R \sim 200 $ GeV, which is well below the typical energy of collisions at the LHC.

\bibliography{LNVNPbib} 

\begin{thebibliography}{10}
\bibitem{ttllBG}
\emph{The \text{SM} cross-section of $p p \to t \bar{t} l^+ l^-$ is estimated
  by multiplying the 38 fb obtained from \texttt{MadGraph5\_aMC@NLO} by the $t
  \bar{t}$ k-factor of 1.6.}, .
\end{thebibliography}
\bibliographystyle{JHEP}

\end{document}